# EFFECT OF INFILL PATTERN AND BUILD ORIENTATION ON MECHANICAL PROPERTIES OF FDM PRINTED PARTS: AN EXPERIMENTAL MODAL ANALYSIS APPROACH

**Santosh Mohan Rajkumar**

Graduate Student
Department of Mechanical & Manufacturing Engineering
Miami University, Oxford, Ohio, USA

**ABSTRACT**

*Fused Deposition Modeling (FDM) is one of the most widely explored additive manufacturing method that uses thermoplastic materials to manufacture products. Mechanical properties of parts manufactured using FDM are influenced by different process parameters involved during manufacturing as they impact the bonding among different layers of cross-section. In this study, the effect of infill patterns and build orientations on the mechanical properties of PLA based parts manufactured using FDM method is studied. Six different infill patterns (triangles, cubic, concentric, tetrahedral, lines, and zigzag) and three different orientations (flatwise, edgewise, and upright) are considered for rectangular beam type parts. For determining mechanical properties of parts with different infill patterns and orientations, flexural bending test is performed. Experimental modal analysis of manufactured parts is performed to observe the relation of natural frequencies with elastic modulus of the parts obtained from flexural bending test. The possibility of experimental modal analysis as an alternative non-destructive method for testing mechanical properties of FDM 3D printed parts is explored.*

**NOMENCLATURE**

| | |
|---|---|
| $\sigma$ | Flexural stress |
| $\varepsilon$ | Flexural strain |
| E | Young's modulus |
| $\Delta$ | slope of load-displacement curve |
| $\omega_n$ | Natural frequency |
| d | Flexural deflection / displacement |
| P | Applied load |
| FDM | Fused Deposition Modeling |
| AM | Additive manufacturing |
| PLA | Polylactic acid |

## 1. INTRODUCTION

There are three fundamental types of manufacturing methods: additive, subtractive, and formative [3]. Out of the three types, additive manufacturing is relatively newer compared to the rest of the two [3]. Additive manufacturing or 3D printing is the process of attaching materials to make objects from 3D models, usually layer by layer, contrary to the traditional subtractive manufacturing methodologies as per ASTM definition [1]. A typical additive manufacturing begins with building of a virtual 3D model in a computer using any Computer Aided Design (CAD) software, then converting the model into a file that defines the geometry of the part to be manufactured (e.g. STL file), thereafter slicing that file into 2-D cross-sections of the part, and then translating the sliced 2D data to a 3D printer or additive manufacturing machine to produce the part with layer by layer deposition of the desired material [1],[3].

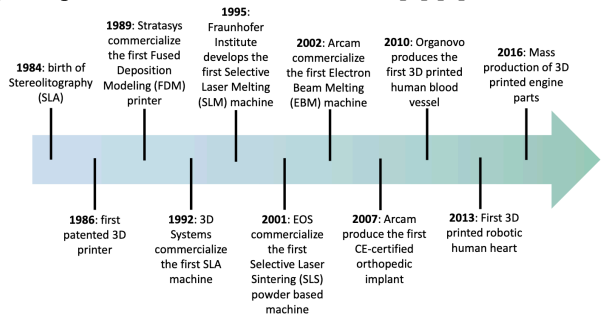

*Figure 1: Historic timeline of additive manufacturing technology [1]*

The major advantages of additive manufacturing are [1]:
- Structures or parts with very complex shape / geometry can be achieved.
- Components can be produced without assembly processes.
- There is no need of component stocking as the manufacturing happens directly from the required material.
- Reduction of waste materials.
- Low consumption of energy compared to traditional manufacturing processes.

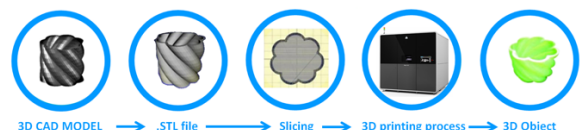

*Figure 2: Steps involved in additive manufacturing process [1]*

In the last few decades, the additive manufacturing technology that has received the most attraction, innovation, and development is Fused Deposition Modeling (FDM) or Fused Filament Fabrication (FFF) [1],[3]. In FDM process, a nozzle moves in a 2-dimensional plan with molten filament of the desired material to deposit each layer of cross-sections of the part to be manufactured on a moveable build platform [3]. The summary of steps involved in FDM AM process are [1]:
- Supply of filament from filament spool to the extrusion head.
- Movement of filament by drive wheels into the nozzle.
- Deposition of filament layer by layer on the build plate with the nozzle.
- Upward movement of the nozzle when a layer is completed.
- Removal of the part once completed.

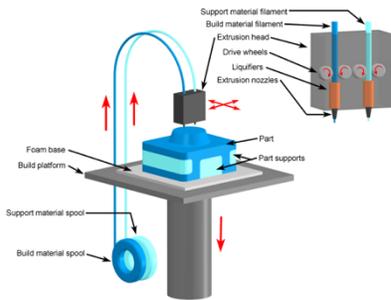

Figure 3: Schematic of FDM AM Process [1]

FDM process is useful for rapid prototyping of thermoplastic parts [2]. It is also simple and low-cost AM process [3]. However, the mechanical strength of parts produced by FDM process are limited by weak bonding of layers and raw filament performances [2]. Bonding quality among deposited filaments determines the mechanical properties and integrity of parts manufactured with FDM process [1]. Bonding is the cohesive force between semi-molten filaments within a layer and between different layers in 3D printed parts [1]. Bonding in FDM AM process is generally of two types [1]:
- **Intra-layer bonding**: bonding between deposited filaments of the same layer.
- **Inter-layer bonding**: bonding between subsequent deposited layers.

FDM AM process finds application in [1]:
- Electrically conductive plastic parts.
- Dies used in sheet metal formation.
- Pharmaceutical industry.
- Tissue engineering.
- Prosthetic structures.
- Dental surgery applications.
- Maxillofacial surgery.
- Ornament industry.
- Electronic circuit manufacturing.

There are various types of materials used in FDM additive manufacturing technology. The most common types of materials used in FDM AM process are [1]:
- PolyCarbonite (PC)
- Acrilonitrile Butadiene Stirene (ABS)
- PolyLactic Acid (PLA)
- Polyetherimide (ULTEM)
- PolyEtherEtherKetone (PEEK)

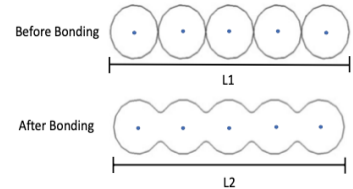

Figure 4: Intra-layer bonding in FDM process [1]

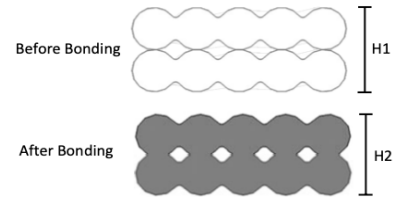

Figure 5: Inter-layer bonding in FDM process [1]

There are various types of materials used in FDM additive manufacturing technology. The most common types of materials used in FDM AM process are [1]:
- PolyCarbonite (PC)
- Acrilonitrile Butadiene Stirene (ABS)
- PolyLactic Acid (PLA)
- Polyetherimide (ULTEM)
- PolyEtherEtherKetone (PEEK)

To improve the mechanical strength / properties of FDM 3D printed parts, high strength filaments can be used, and process parameters can be optimized [3],[2]. Different process parameters involved in FDM AM process are [3],[2],[1],[9]:
- **Infill Density:** It is the fullness of the inside portion of a manufactured part. It can vary from 0 to 100 %. 0% infill implies hollow part, and 100% infill implies solid part. The load bearing capacity of a 3D printed part increases with increase in infill density.
- **Build Orientation:** It is the printing orientation of the part with respect to Z-axis. The X-Y plane is referred to as build platform area and Z-axis refers to height of the printed parts. Build orientation is an important parameter effecting mechanical properties of printed parts
- **Raster Angle:** It is the angle made by the path of the printing nozzle with the x-axis of the build platform. Different studies show the effect of raster angle on the mechanical strength of the printed parts.
- **Layer thickness:** The thickness of each constituent layer of the 2D printed part. Tensile strength is found to be decreasing with increase in layer thickness and compressive strength is found to be increasing with increase in layer thickness.

- **Nozzle diameter:** the air gap between adjacent deposited filament can be controlled using the nozzle diameter and layer thickness during FDM process. With the same layer thickness, increase in nozzle diameter results in higher flexural strength. Increase in nozzle diameter also results in higher tensile strength and stiffness.
- **Extrusion Temperature:** Extrusion temperature can have positive impact on the mechanical strength of 3D printed parts. Especially, for materials like PLA, the fluidity at molten state increases at high temperature facilitating the adhesion between newly deposited filaments with the already existing filaments.
- **Shell thickness:** It is the thickness of the outer shell across the length of the part.

Bellini et al. (2003) [4] showed that the mechanical properties of FDM printed parts depend on build orientations. It has been verified using analytical, numerical, and experimental studies for ABS material. Kovan et al. (2016) [6] studied the effect of layer thickness and print orientation on the mechanical strength of 3D printed parts with PLA. Vishwas et al. (2018) [9] conducted a study to observe the effect of orientation angle, layer thickness, and shell thickness on the ultimate tensile strength of ABS and Nylon materials. Yao et. Al (2019) studies effect of print orientation (angle wise) and layer thickness on mechanical properties of PLA experimentally and theoretically. Beatti et al. (2021) analyzed the effect of print orientation on the mechanical strength of FDM printed ABS parts and explored the failure mode using scanning electron microscopy.

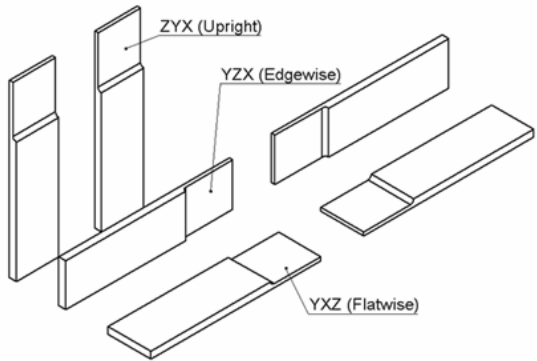

*Figure 6: Print orientations used in [5]*

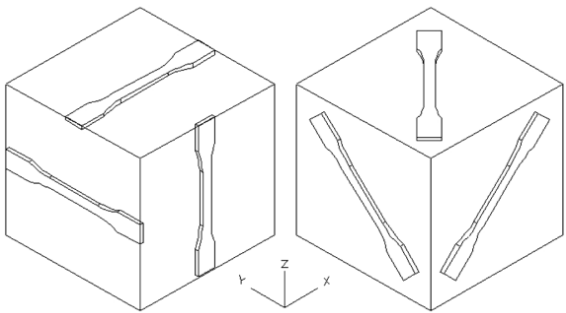

*Figure 7: Print orientations used in [4]*

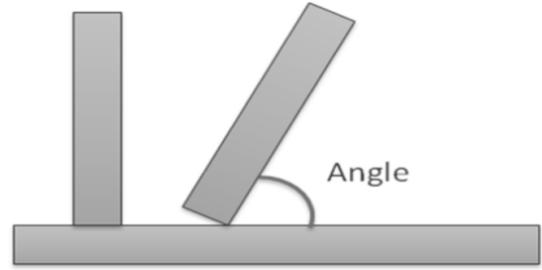

*Figure 8: Print orientations used in [8],[9]*

Chadha et. al (2019) [10] presents the effect of three types of infill patterns (grid, triangular, and honeycomb) on mechanical strength of PLA parts. Akhoundi et. al (2019) [11] studies the effect of different infill patterns and infill percentage on mechanical properties for PLA. The study takes into account of four types of infill patterns: concentric, rectilinear, Hilbert curve, honeycomb. Aloyadi et al. (2020) [12] investigated the effect of four infill (triangle, quarter cubic, tri-hexagon, and grid) patterns on mechanical strength of PLA based FMD manufactured parts using compression tests. [3] suggest that there is a need of further studies on the effect of infill patterns on FDM printed parts as new infill patterns are available.

Kumar et al. (2014) [13] reported that the mechanical properties of a part or sample are directly related with the natural frequencies and vibration mode shapes. The study does not suggest how different mechanical properties are related to the natural frequencies. Osman et al. (2019) [14] investigates the effect of different infill patterns (rectangular, square, triangle, wiggle, criss-cross, honeycomb, full) of beam structures on beam vibration. Kannan et al. (2020) [15] performs experimental modal analysis on three types of materials (ABS, PC, and PC-ABS) for characterization of mechanical properties. The study concludes that the material with better mechanical property has higher natural frequency of vibration. Change of dynamic properties with infill pattern is not adequately addressed in literature [15].

In this study, rectangular beam type structures with six different infill patterns are printed using FDM additive manufacturing process. Each infill pattern is printed with three different print orientations. Five sets of each specimen is printed. The three types of build orientations considered in this study are : flatwise, edgewise, and upright. Also, the six types of infill patterns considered are :
- Lines
- Triangles
- Cubic
- Tetrahedral
- Concentric
- Zig-zag

Each printed sample is subjected to flexural bending test to obtain mechanical properties like the flexural modulus of elasticity, maximum flexural stress, and maximum flexural strain. This allows us to study the effect of different infill patterns and print orientations on mechanical strength of the 3D printed parts. But, flexural test methods are destructive in nature.

Therefore, there is a need to consider non-destructive testing methods. Experimental modal analysis is a powerful non-destructive testing technique to obtain dynamic properties of any part or structure. The relation of natural frequencies of the 3D printed samples obtained from experimental modal analysis with the elastic modulus obtained using flexural testing is compared. The possibility of experimental modal analysis to become an alternative testing method for 3D printed structures for mechanical properties is discussed.

## 2. MATERIALS AND METHODS

The structure of the 3D printed specimen under study is a beam of dimension 100 mm x 15 mm x 5 mm made up of PLA as shown in figure 9. The specimens are printed using Dremel 3D45 3D printer which is an FDM based printer. The six different types of infill patterns considered in this study are shown in figure 10. Different printing orientations for each type of infill pattern have been displayed in figure 11 as seen in the slicing software.

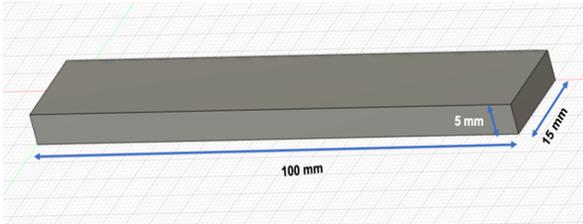

Figure 9: CAD model of the beam specimen with dimensions

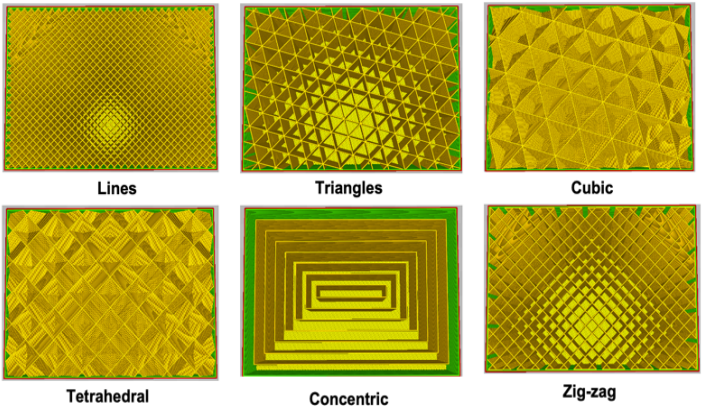

Figure 10: Different infill patterns considered

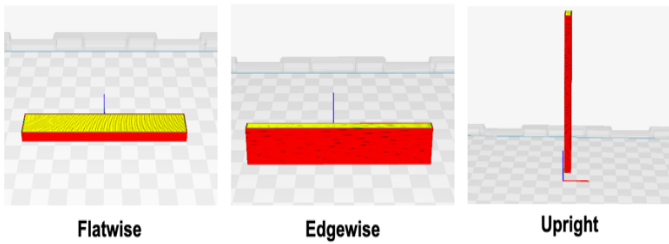

Figure 11: Three print orientations considered in this study

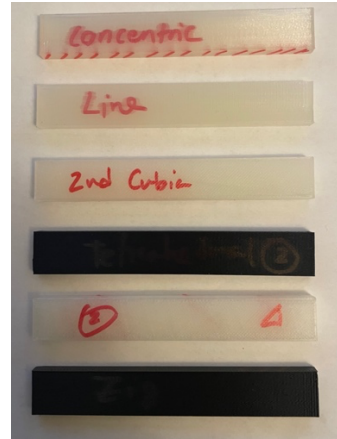

Figure 12: Printed samples in edgewise orientation

### 2.1 Flexural test

Flexural testing of polymeric materials is done as per ASTM standard D 7264/ D 7264M [16]. This test method determines the flexural stiffness and strength properties. There are two ways of doing flexural test:
- Three point bending test
- Four point bending test

Three point bending test is used in this study. In three point bending test, a bar of the test specimen of rectangular cross-section is loaded at a constant rate. Figure 13 Shows the arrangements for three point bending test. The test specimen bar is supported by two supports L distance apart. Distance L is called span of the two supports. Force is applied to the specimen at the center of the span using a loading nose. The force is applied at a constant rate and the deflection due to the applied force is recorded until failure happens or the deformation gets to a predefined value. The mechanical properties obtained using this test method can be utilized for quality control and any design requirement.

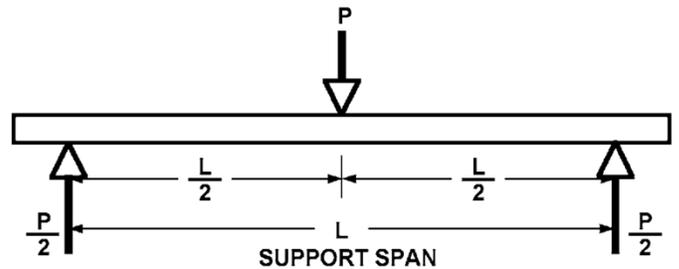

Figure 13: Loading diagram of 3 point bending test [16]

The calculations of mechanical properties using flexural test methods are based on beam theory. The deviation from beam theory increases with increase in width of the specimen. Loading noses used for bending tests may be fixed or rotating. Loading noses need to be in uniform contact with the surface of the specimen across its width. A non-uniform contact may result in non-uniform loading of the specimen and sometimes damage the structure. For uniformity, the loading noses and support structures ideally should possess cylindrical contact surfaces of radius 0.125 in, hardness of 60-62 HRC. The deflection of the specimen at the span center should be measured by a calibrated device with an accuracy of ± 1 % or finer. The deflection needs to be recorded continuously.

Using the load-displacement graph from the test, the stress ($\sigma$) at any point of the curve can be computed using equation (1) where, P is the applied load, w is the thickness of the specimen, t is the thickness of the specimen, and L is the length of span. The maximum flexural stress corresponds to the maximum value of P at the center of the span. The flexural strength of the specimen is equal to the maximum flexural stress at maximum load just before failure or breakdown of the specimen. We can also calculate the flexural strain at each point of load-displacement curve using equation (2) where, d is the displacement or deflection at the center of the span, t is the thickness of the specimen, and L is the length of span.

$$\sigma = \frac{3PL}{2wt^2} \quad (1)$$

$$\varepsilon = \frac{6dt}{L^2} \quad (2)$$

The flexural modulus of elasticity can be defined as the ratio of stress to the respective strain for any point on the stress-strain curve. The flexural modulus of elasticity can be calculated by the formula given in equation (3) using the load-deflection curve from the three-point bending test where, $\Delta$ is the slope of the load-displacement curve and rest of the parameters as defined above.

$$E_f = \frac{\Delta L^3}{4wt^3} \quad (3)$$

Using elastic beam-theory, the deflection or displacement in the three-point loading test is given by equation (4) where, E is the Young's modulus. For a beam of dimension L x w x t, the second moment of inertia I is given as shown in equation (5).

$$d = \frac{PL^3}{48EI} \quad (4)$$

$$I = \frac{1}{12}wt^3 \quad (5)$$

Using equation (5) in (4),

$$E = \frac{PL^3}{d4wt^3} = \frac{\Delta L^3}{4wt^3} \quad (6)$$

Therefore, we can deduct that the flexural modulus obtained from three-point bending test is equal to the Young's modulus in case of rectangular beam structure.

## 2.2 Elastic modulus and natural frequencies

The rectangular beam type 3D printed specimens considered in this study is uniform, slender, and made up of linear elastic material PLA. Therefore, we can use Euler-Bernoulli beam theory of a transversely vibrating beam to derive a relation between the natural frequencies of vibration and Young's modulus as described in [17]. Let us consider a transversely vibrating specimen of length L, rectangular cross section A(x), as shown in figure 14. A distributed force f(x,t) (force/unit length) is acting on the beam. Here, $w(x,t)$ is the deflection in y-direction, bending moment is $M(x,t)$, second-area moment of inertia around the z-axis is $I(x)$, E is the Young's elastic modulus, $\rho$ is the density of the material of the beam. The bending moment acting on the beam is given by equation (7).

$$M(x,t) = EI(x)\frac{\partial^2 w(x,t)}{\partial x^2} \quad (7)$$

If we neglect axial loading, shear deformation, the governing equation of the transverse vibration of a beam under no external force becomes,

$$\rho A(x)\, dx\frac{\partial^2 w(x,t)}{dt^2} + EI\frac{\partial^4 w(x,t)}{\partial x^4} = 0 \quad (8)$$

$$\Rightarrow \frac{\partial^2 w(x,t)}{\partial x^2} + c^2\frac{\partial^4 w(x,t)}{\partial x^4} = 0 \quad (9)$$

Here, w(x,t) is transverse deflection and $c = \sqrt{EI/\rho A}$.

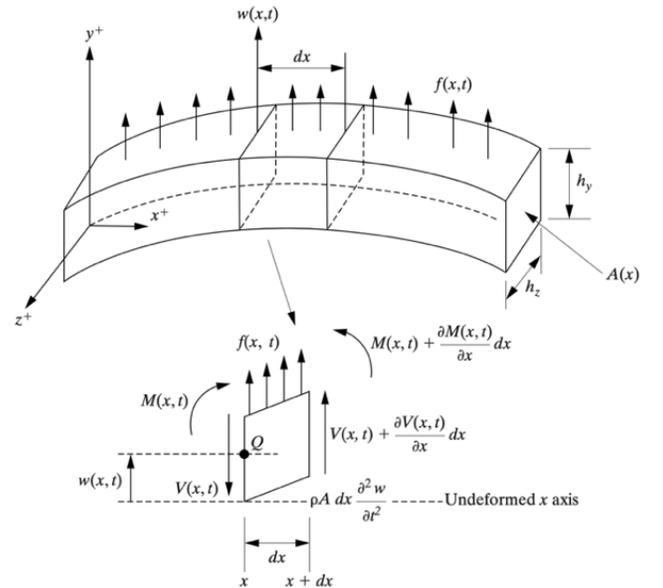

*Figure 14: A transversely vibrating beam and below it the FBD of an infinitesimal element [17]*

Equation (9) is second order in time and fourth order in displacement. Therefore, the solution requires two initial conditions (initial displacement and initial velocity) and four boundary conditions: deflection, slope of deflection, bending moment, and shear force. For brevity, let us consider the scenario where one end of the specimen is fixed, and the other end is free (i.e., cantilever configuration). The boundary conditions for fixed-free configuration are

$$Deflection,\ w(x,t) = 0$$

$$Slope,\ \frac{\partial w(x,t)}{\partial t} = 0 \quad (10)$$

$$M(x,t) = EI\frac{\partial^2 w(x,t)}{\partial x^2} = 0$$

$$V(x,t) = EI\frac{\partial^3 w(x,t)}{\partial x^3} = 0$$

Here, $\beta^4 = \frac{\omega^2}{c^2} = \omega^2 \frac{\rho A}{EI}$. V(x,t) is the transverse shear load.

We assume a solution of (9) in separable form as,
$$w(x,t) = v(x)T(t) \qquad (11)$$

Using (11) in (9), we get,
$$c^2\frac{v''''(x)}{v(x)} = -\frac{\ddot{T}(t)}{T(t)} = \omega^2 (let) \qquad (12)$$

From equation (12) we obtain part of the solution as,
$$v(x) = z_1 \sin\beta x + z_2 \cos\beta x + z_3 \sinh\beta x + z_4 \cosh\beta x \qquad (13)$$

Using the boundary conditions, we arrive at a eigen value problem in matrix form as,

$$\begin{bmatrix} 0 & 1 & 0 & 1 \\ \beta & 0 & \beta & 0 \\ -\beta^2\sin\beta L & -\beta^2\cos\beta L & \beta^2\sinh\beta L & \beta^2\cosh\beta L \\ -\beta^3\cos\beta L & \beta^3\sin\beta L & \beta^3\cosh\beta L & \beta^3\sinh\beta L \end{bmatrix} \begin{bmatrix} z_1 \\ z_2 \\ z_3 \\ z_4 \end{bmatrix} = \begin{bmatrix} 0 \\ 0 \\ 0 \\ 0 \end{bmatrix} \qquad (14)$$

The eigen values can be found by solving the following equation,

$$\begin{vmatrix} 0 & 1 & 0 & 1 \\ \beta & 0 & \beta & 0 \\ -\beta^2\sin\beta L & -\beta^2\cos\beta L & \beta^2\sinh\beta L & \beta^2\cosh\beta L \\ -\beta^3\cos\beta L & \beta^3\sin\beta L & \beta^3\cosh\beta L & \beta^3\sinh\beta L \end{vmatrix} = 0 \qquad (15)$$

$$\Rightarrow \cos\beta L \cosh\beta L + 1 = 0 \qquad (16)$$

The roots of the characteristic equation (16) are found as,

$$\beta_1 L = 1.87510407, \beta_2 L = 4.69409113, \beta_3 L = 7.85475744,$$
$$\beta_4 L = 10.99554073, \beta_5 L = 14.13716839, \beta_n L = \frac{(2n-1)\pi}{2} \ (n>5) \qquad (17)$$

Using the $\beta$ values in equation (17), we can calculate natural frequencies of different modes of the specimen as,

$$\omega_n = \sqrt{\frac{EI}{\rho A}}\beta_n^2 \ (n = 1,2,3 \ldots) \qquad (18)$$

Equation (18) shows that there is a direct relationship of Young's modulus with the natural frequencies of a rectangular beam type sample. Knowing the Young's modulus of the material can provide us the natural frequencies for different modes or vice versa. Therefore, experimental vibration analysis can be done to determine the natural frequencies and comparison of natural frequencies for different infill and/or build orientation can be done to compare the elastic modulus.

## 2.3 Experimental modal analysis

Experimental modal analysis is the method employed to extract modal parameters (natural frequencies, damping, and mode shapes) using time or frequency domain experiments for linear parts or structures [18]. Experimental modal analysis can be performed for simple or complex structures regardless of availability of analytical solutions. For simpler structures like the ones used in this study impact hammer modal testing is used. For impact hammer test, force input is provided to the part under test using the hammer and vibration or acceleration of the part from equilibrium is recorded in time domain. Some of the critical things that need to be taken cared of during impact testing are [2]:

- The hammer tip should be carefully selected. A very hard hammer tip provides a very wider range of frequencies. If very soft tip is used, there is a chance that not all the vibration modes will be excited.
- If the vibration of the specimen does not die out during the sample interval, it may cause a problem to the transducer called leakage. Use of a weighting function called window helps in minimizing leakage. A sample windowing process is shown in figure

The time domain data collected is used to obtain the frequency response function (FRF) by converting the time domain data to frequency domain using fast Fourier transform. Frequency response function (FRF) is the ratio of output response of the structure to the input force in frequency domain. Let us consider a single degree of freedom mass-spring-damper system with input force f(t) given in equation (19).

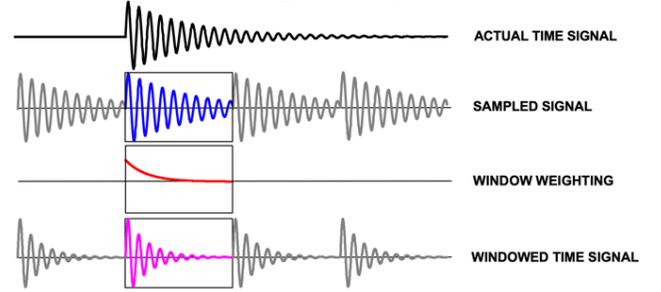

*Figure 15: A sample exponential weighting function [20]*

$$m\ddot{x} + c\dot{x} + kx = f(t) \qquad (19)$$

Taking Laplace transform of (19),
$$(s^2 m + sc + k)X(s) = F(s) \qquad (20)$$
$$\Rightarrow H(s) = \frac{X(s)}{F(s)} = \frac{1}{s^2 m + sc + k} \qquad (21)$$
$$H(j\omega) = H(s)|_{s=j\omega} = \frac{1}{-m\omega^2 + jc\omega + k} \qquad (22)$$

Equation (22) gives the FRF for the single degree of freedom system. For multiple degrees of freedom systems, the FRF is the sum of all individual degree of freedom's FRFs. FRF is the most important measured data required for experimental modal analysis [20]. Computing environments like MATLAB offers functionality to generate FRFs from experimental data obtained from tests like hammer test [19]. Once we obtain the FRF, modal parameters like natural frequencies, damping, and mode shapes can be extracted. First the FRF is decomposed into several single degree of freedom systems as shown in figure 16. Then mathematical algorithms for curvefitting estimates the modal parameters. MATLAB signal processing toolbox provides modal fitting functionality using least squares complex exponential method.

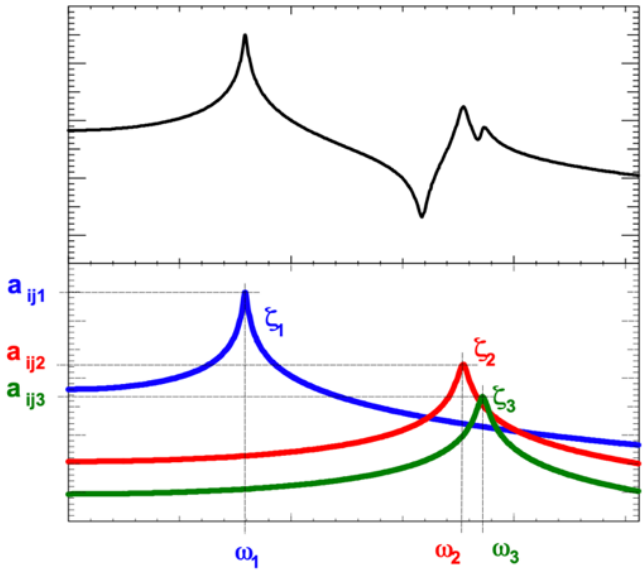

Figure 16: Showing decomposition of an FRF to multiple single degree of freedom systems [20]

### 3. Experimental Results

The specimens with six different infill patterns as shown in figure 10 are considered and each infill pattern is printed with the three different orientations shown in figure 11. Rest of the process parameters for the printing process kept same for all the samples. Other process parameters used during FDM manufacturing process are given in table 1. Five sets of each specimen are printed for testing of mechanical strength.

Table 1: Different process parameter values used for printing the specimens

| FDM Process Parameter | Value |
|---|---|
| Nozzle diameter | 0.4 mm |
| Infill density | 100 % |
| Layer height | 0.2 mm |
| Shell thickness | 0.8 mm |
| Extrusion temperature | $230^o\ C$ |
| Build plate temperature | $60^o\ C$ |
| Filament diameter | 1.75 mm |

The printed samples are first tested for mechanical strength using three point bending test. The test is performed using Instron 5867. Each sample specimen is kept on the span of length L= 2.45 inch and the force is applied at the mid of the span at a uniform rate of 0.5 inch per minute. The sampling time of collected data is 0.015 s. The applied load has a force sensor which gives reading in lbf. The force-deflection data is obtained in the computer connected to the testing machine.

The force-displacement data obtained from the Instron 5867 test machine are converted into SI units. The plots of force-displacement curve of one sample for each type of infill pattern with all the three orientations are shown in figures 19-24.

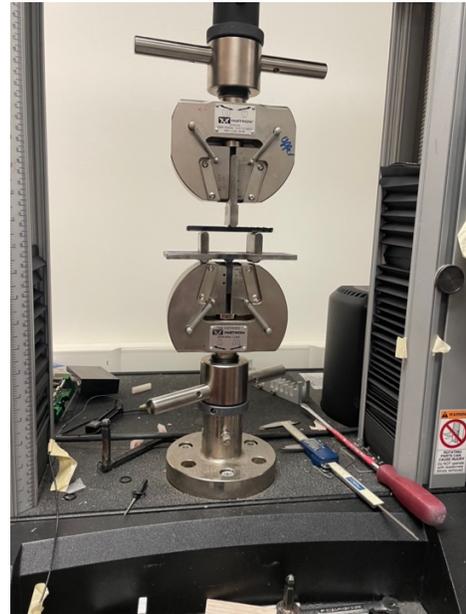

Figure 17: Instron 5867 test machine for three point bending test

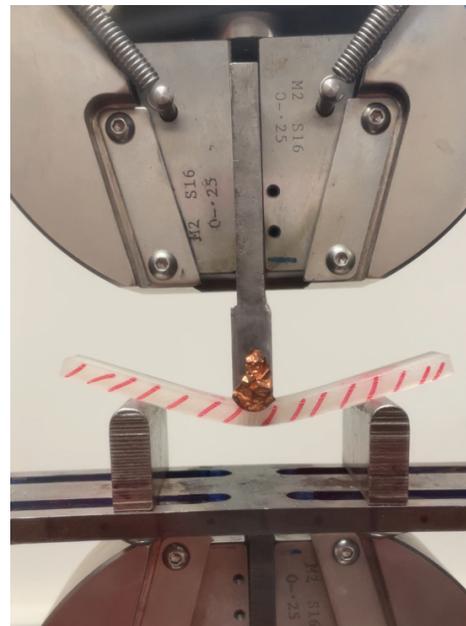

Figure 18: A test specimen undergoing three point bending test

For each point of load-displacement curve of each specimen, corresponding flexural stress and flexural strain is calculated using equations (1) and (2). Then maximum flexural stresses and flexural strains are obtained at the points where load is maximum. Average of the maximum flexural stresses and flexural strains of the five samples of each infill pattern with each print orientation are considered as the representative of that pattern and orientation. For determining the flexural modulus of elasticity or in this case Young's modulus (as the specimens are beam type), we find the slope of the load-displacement curve in the linear elastic range much before the breakdown point. Then the formula given in equation (3) is used to calculate the flexural modulus of each specimen. The average Young's modulus of the five samples of each infill pattern with each print orientation is considered to be the representative of that pattern and orientation. Summary of mechanical properties of each infill pattern with each orientation is shown in table 2.

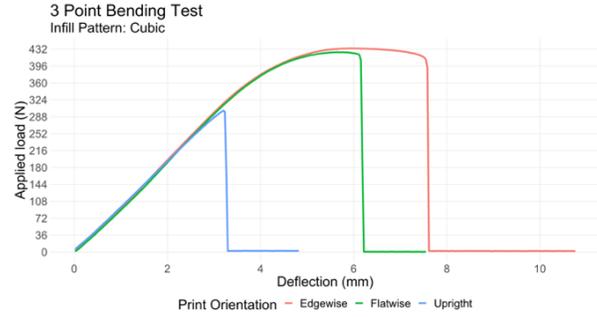

Figure 22: Load-displacement curve for three samples with cubic pattern and different orientations

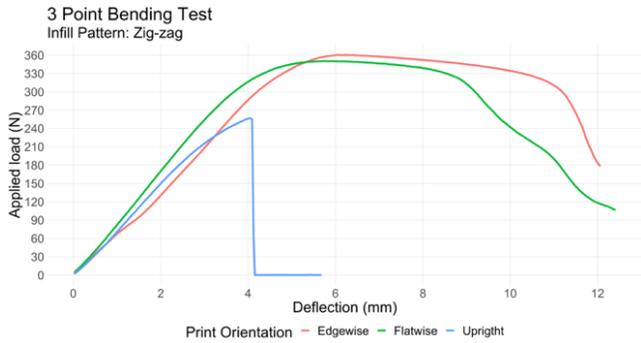

Figure 19: Load-displacement curve for three samples with zigzag pattern and different orientation

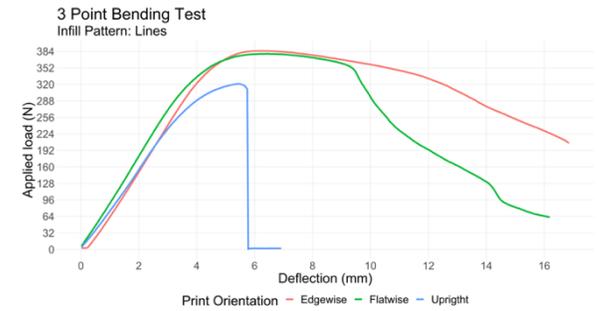

Figure 23: Load-displacement curve for three samples with lines pattern and different orientations

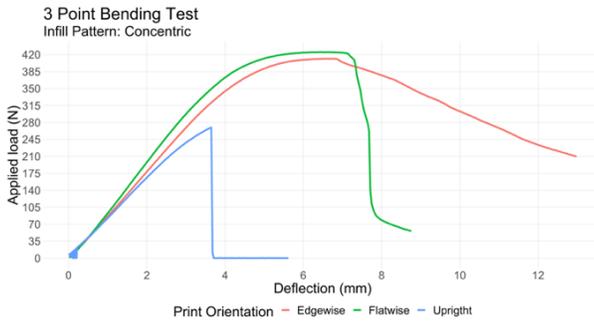

Figure 20: Load-displacement curve for three samples with concentric pattern and different orientations

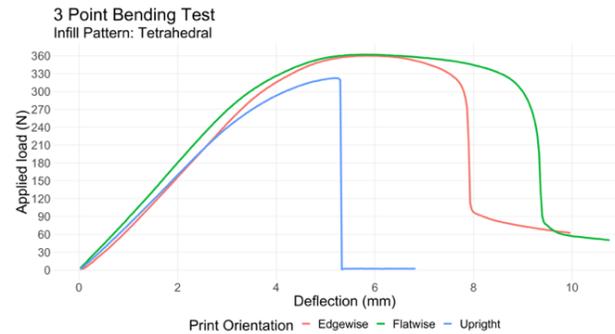

Figure 24: Load-displacement curve for three samples with tetrahedral pattern and different orientations

Figure 25 shows the maximum flexural strength of the specimens. We observe that cubic infill pattern with edgewise orientation has the highest maximum flexural stress of 107.95 MPa, closely followed by concentric pattern with flatwise orientation and cubic pattern with flatwise orientation. Triangles pattern with upright orientation has the lowest maximum flexural stress of 62.32 MPa closely followed by zigzag pattern with upright orientation. With flatwise print orientation, cubic infill pattern has the highest maximum flexural stress (closely followed by concentric pattern) and zigzag infill pattern has the lowest maximum flexural stress. With edgewise orientation, cubic infill pattern has the highest flexural stress as already mentioned and tetrahedral infill pattern has the lowest maximum flexural stress (closely followed by zigzag infill pattern). With Upright print orientation, tetrahedral infill pattern has the highest maximum flexural stress (closely followed by lines infill pattern) and triangles infill pattern has the lowest maximum flexural stress. Figure 26 shows boxplot of different infill patterns with maximum infill pattern regardless of print orientations. We can clearly see that cubic patterns tend to have higher maximum flexural strength compared to other infill patterns. Zigzag patterns tend to have lower maximum flexural strength compared to other infill patterns. From figure 27, we observe that median value of maximum flexural stress in flatwise orientation is higher than rest of the two orientations. In general, flatwise orientations tend to have higher maximum flexural stress. Upright print orientation tend to have much lower maximum flexural stress compared to the other two types.

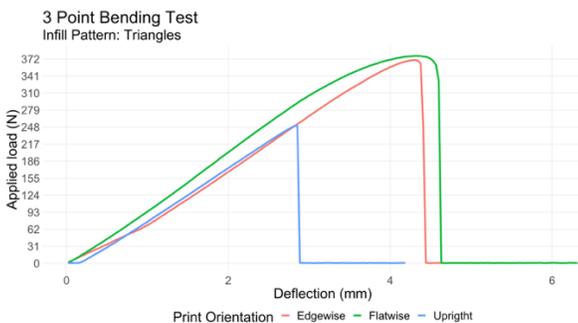

Figure 21: Load-displacement curve for three samples with triangles pattern and different orientations

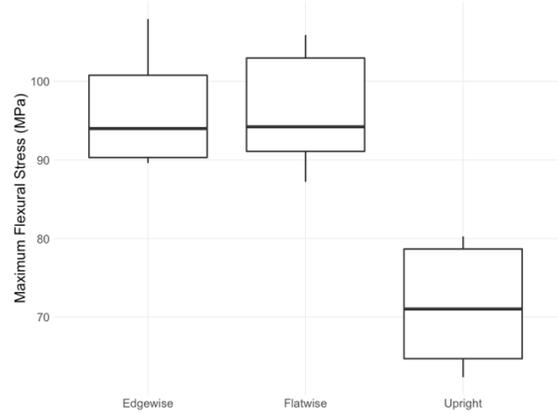

*Figure 27: Boxplot of different print orientations with maximum flexural stress regardless of infill patterns*

Figure 28 shows the flexural modulus obtained for different infill patterns and orientations. We observe that triangles infill pattern with flatwise orientation has the highest flexural modulus of 3.36 GPa and zigzag infill pattern with edgewise orientation has the lowest flexural modulus of 2.185 GPa. With flatwise orientation, triangles infill pattern has the highest flexural modulus and zigzag infill pattern has the lowest flexural modulus. With edgewise orientation, cubic infill pattern has the highest flexural modulus and zigzag infill pattern has the lowest flexural modulus. With upright orientation, cubic infill pattern has the highest flexural modulus (closely followed by triangles infill pattern) and zigzag infill pattern has the lowest flexural modulus.

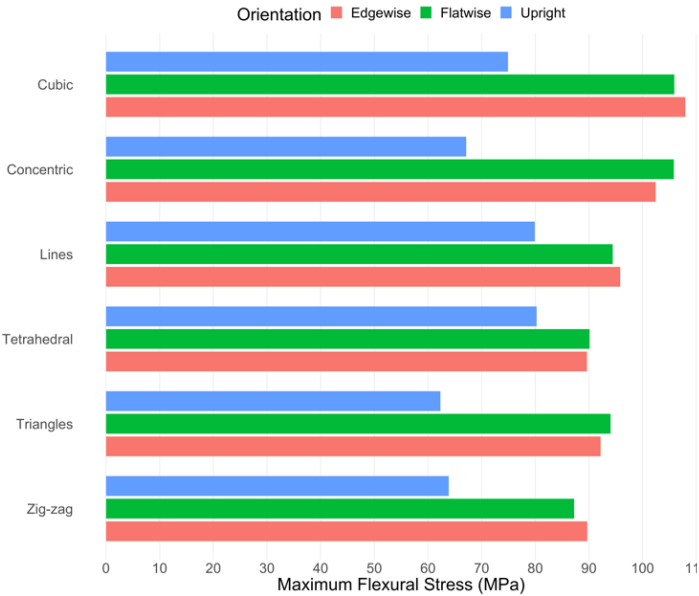

*Figure 25: Bar graph showing maximum flexural stress of different infill patterns with different orientations*

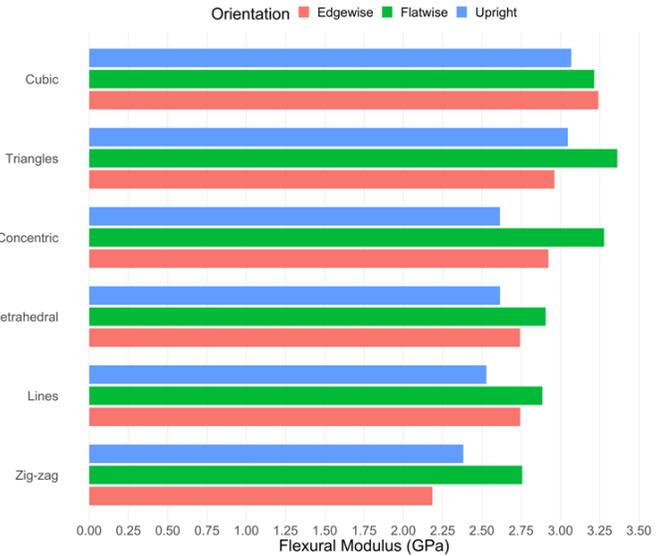

*Figure 28: Bar graph showing flexural modulus with different infill patterns and print orientations*

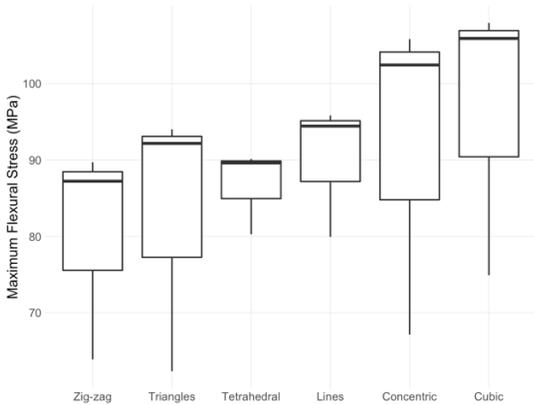

*Figure 26: Boxplot of different infill patterns with maximum flexural stress regardless of print orientations*

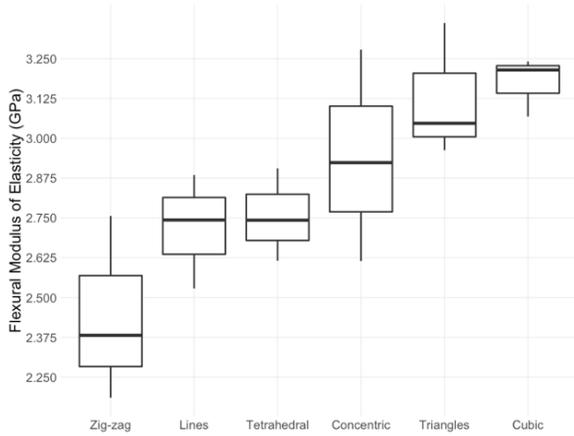

*Figure 29: Boxplot of different infill patterns with flexural modulus regardless of print orientations*

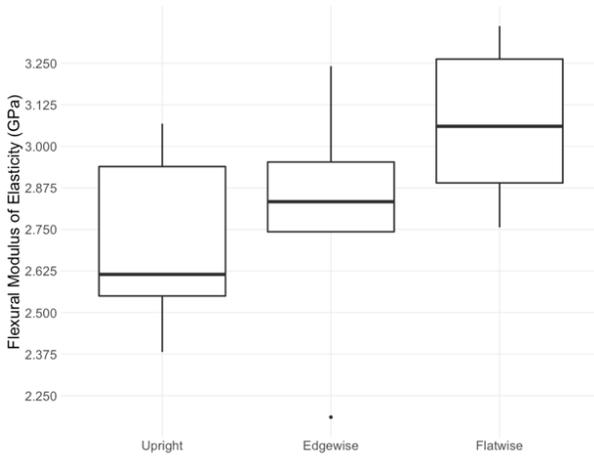

*Figure 30: Boxplot of different print orientations with flexural modulus regardless of infill patterns*

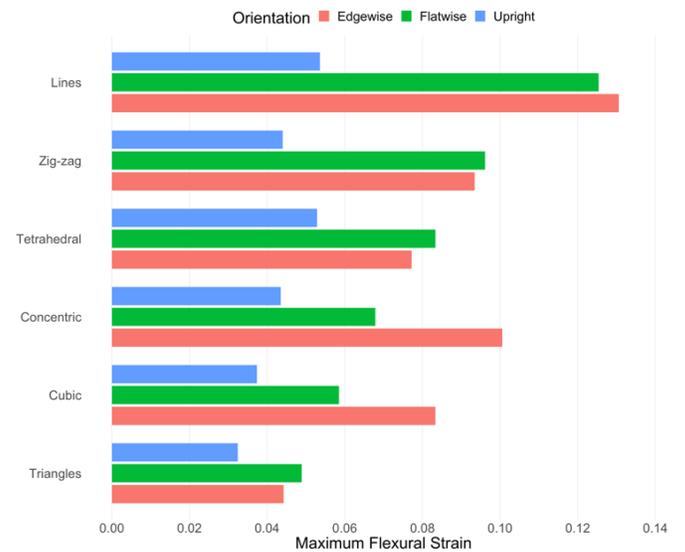

*Figure 31: Bar graph showing maximum flexural strain with different infill patterns and print orientations*

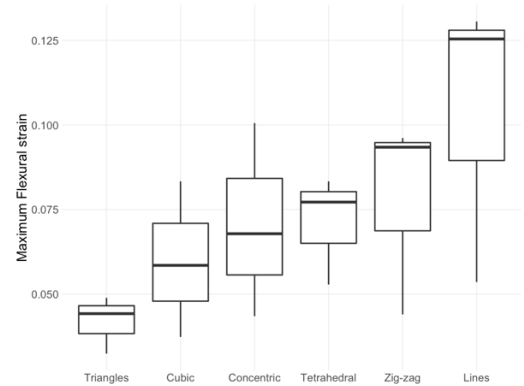

*Figure 32: Boxplot of different infill patterns with maximum flexural strain regardless of print orientations*

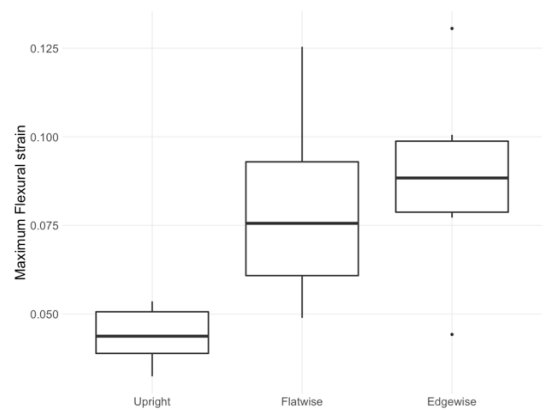

*Figure 33: Boxplot of different print orientations with maximum flexural strain regardless of infill patterns*

Figure 29 shows boxplot of flexural modulus for all the infill patterns regardless of print orientation. We observe that cubic infill pattern tends to have higher flexural modulus compared to other patterns and zigzag pattern tends to have lower flexural modulus compared to other infill patterns. Figure 30 shows boxplot of flexural modulus for all the print orientations regardless of infill pattern. Flatwise orientation tends to have higher flexural modulus and upright orientation tend to have lower flexural modulus compared to other two orientations.

Figure 31 shows the bar graph of maximum flexural strain with all infill patterns and orientations. We observe that lines infill pattern with edgewise orientation has the highest maximum flexural strain and triangles infill pattern with upright orientation has the lowest maximum flexural strain. With flatwise orientation, lines infill pattern has the highest and triangles infill pattern has the lowest maximum flexural strain. With edgewise orientation, similar pattern is observed. With upright orientation, tetrahedra pattern has the highest and triangles pattern has the lowest maximum flexural strain.

From figure 32, we observe that lines infill pattern tends to have higher and triangles infill pattern tends to have lower maximum flexural strain. Applications where ductile property is desired lines pattern is a plausibly better choice over the other patterns considered in this study. From figure 33, we observe that edgewise orientation tends to have higher maximum flexural strain over the other two. Applications where ductile property is desired edgewise orientation is a plausibly better choice over the other orientations considered in this study.

*Table 2: Summary of results obtained from three point bending test*

| Infill Pattern | Print Orientation | Flexural Modulus (GPa) | Maximum Flexural Stress (MPa) | Maximum Flexural Strain |
|---|---|---|---|---|
| Zig-zag | Flatwise | 2.756254 | 87.21 | 0.0961353 |
| Zig-zag | Edgewise | 2.185246 | 89.69 | 0.0934298 |
| Zig-zag | Upright | 2.381209 | 63.87 | 0.0439916 |
| Concentric | Flatwise | 3.278420 | 105.83 | 0.0678498 |
| Concentric | Edgewise | 2.923561 | 102.44 | 0.1005626 |
| Concentric | Upright | 2.614269 | 67.13 | 0.0434957 |
| Triangles | Flatwise | 3.362011 | 94.01 | 0.0489108 |
| Triangles | Edgewise | 2.962456 | 92.16 | 0.0442356 |
| Triangles | Upright | 3.047025 | 62.32 | 0.0324295 |
| Cubic | Flatwise | 3.214655 | 105.91 | 0.0585033 |
| Cubic | Edgewise | 3.241197 | 107.95 | 0.0833454 |
| Cubic | Upright | 3.068235 | 74.90 | 0.0373506 |
| Tetrahedral | Flatwise | 2.905575 | 90.13 | 0.0833454 |
| Tetrahedral | Edgewise | 2.742876 | 89.61 | 0.0771963 |
| Tetrahedral | Upright | 2.615264 | 80.26 | 0.0528462 |
| Lines | Flatwise | 2.884770 | 94.43 | 0.1254067 |
| Lines | Edgewise | 2.743377 | 95.83 | 0.1305699 |
| Lines | Upright | 2.528250 | 79.92 | 0.0535841 |

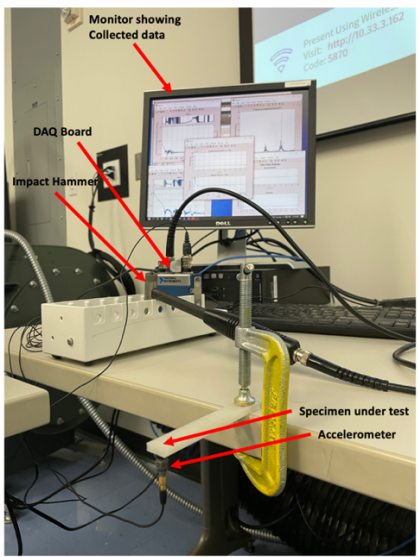

*Figure 34: Test set up experimental modal analysis*

Following are the equipment/tools used for the experimental modal analysis experiment:
- Impact hammer: The impact hammer used is PCB 086C03. The sensitivity range of the hammer is 0-500 lbf with a sensitivity of 2.25 mv/N (10 mV/lbf) (±15 %). The mass of the hammer is 0.16 kg. It comes with various soft & hard tips.
- Accelerometer: The accelerometer used is PCB 352C66 which uses ceramic sensing element. The weight of the accelerometer is 2 g. The sensitivity is 100 mV/g (±10 %). It can measure up to ± 50 g peak acceleration with a resolution of 0.00016 g rms. The frequency of vibration it can sense is 0.5-10000 Hz (±5 %).
- Data Acquisition System (DAQ): The DAQ system is used for converting the analog output signals. NI cDAQ 9271 board with a 9233 module. The sampling frequency of acquired data is 1000 Hz.

The simplified steps involved in experimental modal analysis are:
- Each sample is clamped to the test bench with fixed-free configuration.
- The accelerometer is attached to the tip of the free end.
- Using the impact hammer, force is applied like an impulse and let the specimen vibrate freely.
- The force-acceleration data is acquired in terms of voltage in the computer using MATLAB script.
- In modal analysis we normalize the FRF value, hence conversion of voltage values to forces and acceleration is not required.
- MATLAB's *modalfrf* function is used to generate FR for the experimental data.
- Then modal curve fitting is done on the FRF obtained to get natural frequencies of each specimen.
- First five vibration modes are considered.

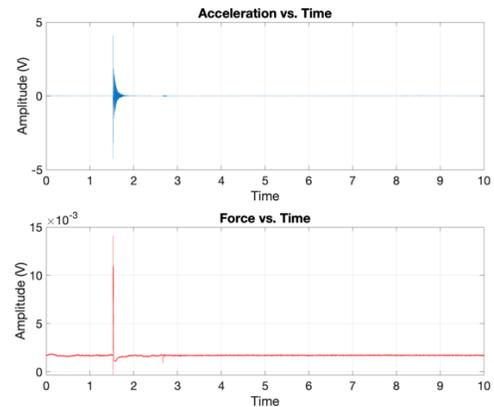

*Figure 35: A sample vibration data for one of the specimens*

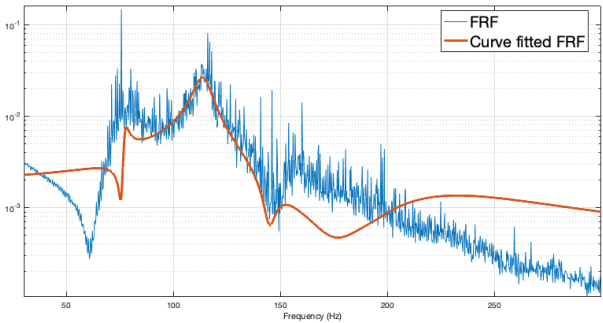

*Figure 36: A sample FRF and modal curve fitted FRF for one of the specimens*

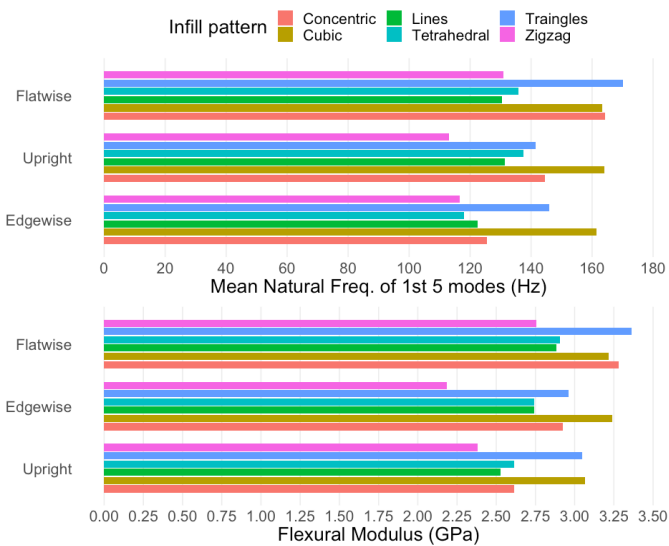

*Figure 37: Comparison of bar charts of natural frequencies and flexural modulus of different infill patterns and print orientations.*

the natural frequency of the specimens and showed relationship with modulus of elasticity. Other dynamic properties like mode shapes, damping ratio may impact the other mechanical properties of the material. Natural frequencies might also impact other mechanical properties. If these facts are addressed properly, we can have experimental modal analysis as a proven testing method for mechanical testing of 3D printed parts. [15] also compared the first five natural frequencies obtained from experimental modal analysis for three materials ABS, PC, PC-ABS (in increasing order of mechanical strength). The natural frequencies with fixed-free condition for first five natural frequencies found to be in increasing order as seen in figure 40.

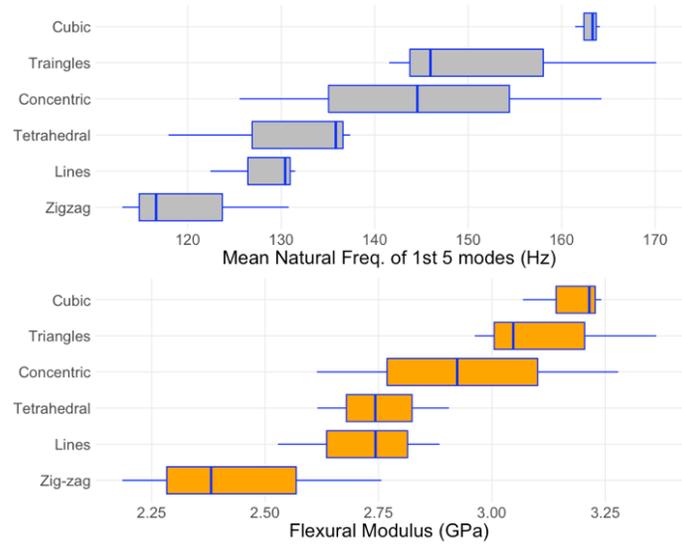

*Figure 38: Comparison of natural frequencies and flexural modulus of different infill patterns regardless of print orientation*

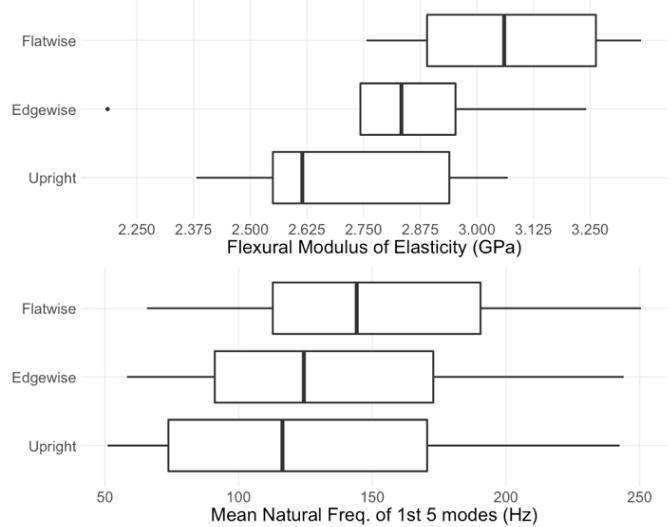

*Figure 39: Comparison of natural frequencies and flexural modulus of different print orientations regardless of infill patterns*

For each specimen, first five natural frequencies are extracted using MATLAB from experimental data. The mean of the first five natural frequencies is taken for comparison. From figure 37, we observe that the mean natural frequency of the first five modes varies in the same way as flexural modulus does for all the patterns and orientations. This finding is in unison with the theoretical finding of beam theory. Figure 38 and 39 shows comparison of box plots of natural frequencies and flexural modulus for infill patterns only and infill orientations only. We observe that in both cases, mean natural frequency of the first five modes varies in the same pattern as that of flexural modulus.

In this study, we have confirmed the proportional relation of natural frequencies with flexural modulus (or modulus of elasticity) for FDM 3D printed parts. Equation (18) gives relation between modulus of elasticity and natural frequency of a particular mode. Therefore, using density of the specimen modulus of elasticity of it can be determined. This corroborates the fact that experimental modal analysis can be an alternative non-destructive testing method for finding mechanical properties of 3D printed materials. In this study, we have only considered

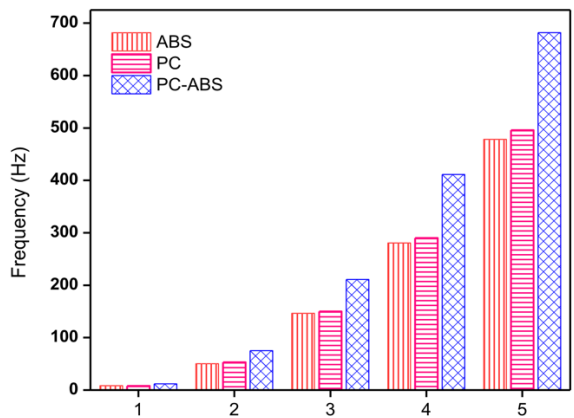

*Figure 40: First five natural frequencies obtained for three types of materials with increasing mechanical strength in [15]*

## 4. CONCLUSION

In this study we have observed the impact of infill pattern and build orientation on the mechanical properties of FDM printed parts. We observe that mechanical properties like modulus of elasticity, maximum flexural stress, maximum flexural strain changes for different infill pattern and orientations. Among the six infill patterns considered in this study, cubic patterns tend to have the highest modulus of elasticity closely followed by triangles pattern and zigzag infill patterns tend to have the lowest flexural modulus. Also, cubic patterns tend to have the highest maximum flexural strength closely followed by concentric pattern and zigzag infill patterns tend to have the lowest maximum flexural strength. Lines infill patterns tend to have the highest maximum flexural strain and triangles patterns tend to have the lowest maximum flexural strain. Among the three print orientations, flatwise orientation tends to have higher modulus of elasticity. Flatwise orientation also tends to have higher maximum flexural stress closely followed by edgewise orientation. Edgewise orientation tends to have higher maximum flexural strain compared to the other two types. Experimental modal analysis of the 3D printed specimens reveals the direct relation of natural frequencies with elastic modulus. Samples with higher elastic modulus tend to have higher natural frequencies. This finding establishes the possibility of experimental modal analysis to be a non-destructive alternative to other conventional mechanical testing methods for 3D printed materials. We have not explored other dynamic properties like mode shapes, damping ratio with other mechanical properties. Future scope of this study can be of relating other dynamic properties with different mechanical properties of 3D printed parts to concretely establish experimental modal analysis as a powerful mechanical testing method.

**REFERENCES**

[1] Cattenone A. ,Analysis and Simulation of Additive Manufacturing Process, *PhD Dissertation, Universita Degli Studi Di Pavia*, 2018 (Link)

[2] Peng, W.A.N.G., Bin, Z.O.U., Shouling, D.I.N.G., Lei, L.I. and Huang, C., 2021. Effects of FDM-3D printing parameters on mechanical properties and microstructure of CF/PEEK and GF/PEEK. *Chinese Journal of Aeronautics,* 34(9), pp.236-246.

[3] Syrlybayev, D., Zharylkassyn, B., Seisekulova, A., Akhmetov, M., Perveen, A. and Talamona, D., 2021. *Optimisation of Strength Properties of FDM Printed Parts*—A Critical Review. Polymers, 13(10), p.1587.

[4] Bellini, A. and Güçeri, S., 2003. Mechanical characterization of parts fabricated using fused deposition modeling. *Rapid Prototyping Journal.*

[5] Kovan, V., Altan, G. and Topal, E.S., 2017. Effect of layer thickness and print orientation on strength of 3D printed and adhesively bonded single lap joints. *Journal of Mechanical Science and Technology*, 31(5), pp.2197-2201.

[6] Wu, W., Ye, W., Wu, Z., Geng, P., Wang, Y. and Zhao, J., 2017. Influence of layer thickness, raster angle, deformation temperature and recovery temperature on the shape-memory effect of 3D-printed polylactic acid samples. *Materials*, 10(8), p.970.

[7] Yao, T., Deng, Z., Zhang, K. and Li, S., 2019. A method to predict the ultimate tensile strength of 3D printing polylactic acid (PLA) materials with different printing orientations. *Composites Part B: Engineering*, 163, pp.393-402

[8] Beattie, N., Bock, N., Anderson, T., Edgeworth, T., Kloss, T. and Swanson, J., 2021. Effects of Build Orientation on Mechanical Properties of Fused Deposition Modeling Parts. *Journal of Materials Engineering and Performance*, pp.1-7.

[9] Vishwas, M., Basavaraj, C.K. and Vinyas, M., 2018. Experimental investigation using taguchi method to optimize process parameters of fused deposition Modeling for ABS and nylon materials. *Materials Today: Proceedings*, 5(2), pp.7106-7114.

[10] Chadha, A., Haq, M.I.U., Raina, A., Singh, R.R., Penumarti, N.B. and Bishnoi, M.S., 2019. Effect of fused deposition modelling process parameters on mechanical properties of 3D printed parts. World Journal of Engineering.

[11] Akhoundi, B. and Behravesh, A.H., 2019. Effect of filling pattern on the tensile and flexural mechanical properties of FDM 3D printed products. Experimental Mechanics, 59(6), pp.883-897.

[12] Aloyaydi, B., Sivasankaran, S. and Mustafa, A., 2020. Investigation of infill-patterns on mechanical response of 3D printed poly-lactic-acid. Polymer Testing, 87, p.106557.

[13] Kumar, A., Jaiswal, H., Jain, R. and Patil, P.P., 2014. Free vibration and material mechanical properties influence based frequency and mode shape analysis of transmission gearbox casing. Procedia Engineering, 97, pp.1097-1106

[14] İYİBİLGİN, O., DAL, H., GEPEK, E.,EXPERIMENTAL MODAL ANALYSIS OF 3D PRINTED BEAMS, 4th INTERNATIONAL CONGRESS ON 3D PRINTING (ADDITIVE MANUFACTURING) TECHNOLOGIES AND DIGITAL INDUSTRY, 2019

[15] Kannan, S. and Ramamoorthy, M., 2020. Mechanical characterization and experimental modal analysis of 3D Printed ABS, PC and PC-ABS materials. Materials Research Express, 7(1), p.015341

[16] ASTM International, 2015. Standard test method for flexural properties of polymer matrix composite materials. ASTM International.

[17] Rao, S.S., 2019. Vibration of continuous systems. John Wiley & Sons.

[18] Inman, D.J. and Singh, R.C., 1994. Engineering vibration (Vol. 3). Englewood Cliffs, NJ: Prentice Hall.

[19] MathWorks, Inc, 2002. Signal Processing Toolbox for Use with MATLAB: User's Guide. The MathWorks.

[20] Avitabile, P., 2001. Experimental modal analysis. Sound and vibration, 35(1), pp.20-31.